\documentstyle[12pt]{article} 
\input psfig.sty

\def\Title#1{\begin{center} {\Large #1 } \end{center}}
\def\Author#1{\begin{center}{ \sc #1} \end{center}}
\def\Address#1{\begin{center}{ \it #1} \end{center}}

\def\doeack{\footnote{Work supported by the Department of Energy,
                     contract DE--AC03--76SF00515.}}
\def\SLAC{Stanford Linear Accelerator Center\\
    Stanford University, Stanford, California 94309 USA}

\newenvironment{Abstract}{\begin{quotation} \begin{center}
                       ABSTRACT
     \end{center}\bigskip  }{\end{quotation}}
\def\beq{\begin{equation}}
\def\eeq#1{\label{#1}\end{equation}}
\def\eeqn{\end{equation}}
\def\beqa{\begin{eqnarray}}
\def\eeqa#1{\label{#1}\end{eqnarray}}
\def\eeqan{\end{eqnarray}}

\def\Acknowledgements{\bigskip  \bigskip \begin{center} \begin{large}
             \bf ACKNOWLEDGEMENTS \end{large}\end{center}}
\def\Re{{\cal R \mskip-4mu \lower.1ex \hbox{\it e}\,}}
\def\Im{{\cal I \mskip-5mu \lower.1ex \hbox{\it m}\,}}
\def\nn{\noindent}
\def\ie{{\it i.e.}}
\def\eg{{\it e.g.}}

\def\etal{{\it et al.}}

\def\sub#1{_{\lower.25ex\hbox{$\scriptstyle#1$}}}
\def\sul#1{_{\kern-.1em#1}}
\def\sll#1{_{\kern-.2em#1}}  
\def\sbl#1{_{\kern-.1em\lower.25ex\hbox{$\scriptstyle#1$}}}
\def\ssb#1{_{\lower.25ex\hbox{$\scriptscriptstyle#1$}}}
\def\sbb#1{_{\lower.4ex\hbox{$\scriptstyle#1$}}}

\def\to{\rightarrow}
\def\dk{\ifmmode \Delta\kappa\else $\Delta\kappa$\fi}
\def\sigt{\ifmmode \tilde\sigma\else $\tilde\sigma$\fi}
\def\mh{\ifmmode m\sbl H \else $m\sbl H$\fi}
\def\mch{\ifmmode m_{H^\pm} \else $m_{H^\pm}$\fi}
\def\mt{\ifmmode m_t\else $m_t$\fi}
\def\mc{\ifmmode m_c\else $m_c$\fi}
\def\mz{\ifmmode M_Z\else $M_Z$\fi}
\def\mw{\ifmmode M_W\else $M_W$\fi}
\def\mws{\ifmmode M_W^2 \else $M_W^2$\fi}
\def\mhs{\ifmmode m_H^2 \else $m_H^2$\fi}   
\def\mzs{\ifmmode M_Z^2 \else $M_Z^2$\fi}
\def\mts{\ifmmode m_t^2 \else $m_t^2$\fi}
\def\mcs{\ifmmode m_c^2 \else $m_c^2$\fi}
\def\mchs{\ifmmode m_{H^\pm}^2 \else $m_{H^\pm}^2$\fi}
\def\ztwo{\ifmmode Z_2\else $Z_2$\fi}
\def\zone{\ifmmode Z_1\else $Z_1$\fi}
\def\mtwo{\ifmmode M_2\else $M_2$\fi}
\def\mone{\ifmmode M_1\else $M_1$\fi}
\def\tb{\ifmmode \tan\beta \else $\tan\beta$\fi}
\def\xw{\ifmmode x\sub w\else $x\sub w$\fi}
\def\ch{\ifmmode H^\pm \else $H^\pm$\fi}
\def\lum{\ifmmode {\cal L}\else ${\cal L}$\fi}
\def\inpb{\ifmmode {\rm pb}^{-1}\else ${\rm pb}^{-1}$\fi}
\def\infb{\ifmmode {\rm fb}^{-1}\else ${\rm fb}^{-1}$\fi}
\def\epem{\ifmmode e^+e^-\else $e^+e^-$\fi}
\def\ppb{\ifmmode \bar pp\else $\bar pp$\fi}

\def\bsg{\ifmmode b\rightarrow s\gamma \else $b\rightarrow s\gamma$\fi}
 
\newskip\zatskip \zatskip=0pt plus0pt minus0pt
\def\matth{\mathsurround=0pt}

\def\atversim#1#2{\lower0.7ex\vbox{\baselineskip\zatskip\lineskip\zatskip
  \lineskiplimit 0pt\ialign{$\matth#1\hfil##\hfil$\crcr#2\crcr\sim\crcr}}}

\begin{document}
\rightline{\vbox{\halign{&#\hfil\cr
&SLAC-PUB-7155\cr
&May 1996\cr}}}
\vspace{0.8in} 
\Title{The Polarization Asymmetry and Triple Gauge Boson Couplings in 
$\gamma e$ Collisions at the NLC
%\footnote{To appear in {\it Physics and Technology of the Next Linear 
%Collider}, eds. D.\ Burke and M.\ Peskin, reports submitted to Snowmass 1996}
}
\bigskip
\Author{Thomas G. Rizzo\doeack}
\Address{\SLAC}
\bigskip
\begin{Abstract}
 
We examine the capability of the NLC in the $\gamma e$ collider mode to probe 
the CP-conserving $\gamma WW$ and $\gamma ZZ$ anomalous couplings through the 
use of the polarization asymmetry. When combined with other measurements, very 
strong constraints on both varieties of anomalous couplings can be obtained. 

\end{Abstract}
\bigskip

\vskip1.0in
\begin{center}
To appear in {\it Physics and Technology of the Next Linear 
Collider}, eds. D.\ Burke and M.\ Peskin, reports submitted to Snowmass 1996.
\end{center}
\bigskip

\def\thefootnote{\fnsymbol{footnote}}
\setcounter{footnote}{0}
\newpage
\section{Introduction}

The Standard Model(SM) has so far done an excellent job at describing almost 
all existing data. One of the most crucial remaining set of tests of the gauge 
structure of the SM will occur at future colliders when precision measurements 
of the various triple gauge boson vertices(TGVs) become 
available{\cite {rev}}. If new physics arises at or near the TeV scale, 
then on rather general grounds one expects 
that the deviation of the TGVs from their canonical SM values, \ie, 
the anomalous 
couplings, to be {\it at most} ${\cal O}(10^{-3}-10^{-2})$ with the smaller 
end of this range of values being the most likely. To get to 
this level of precision, and beyond, for all of the TGVs a number of 
different reactions 
need to be studied using a variety of observables. Here we concentrate on the 
CP-conserving $\gamma WW$ and $\gamma ZZ$ anomalous couplings that can be 
probed in the reactions $\gamma e \to W\nu ,Ze$ at the NLC using polarized 
electrons and polarized 
backscattered laser photons{\cite {old}}. In the $\gamma WW$ case, the 
anomalous 
couplings modify the magnitude and structure of the already existing SM tree 
level vertex. No corresponding tree level $\gamma ZZ$ vertex exists in 
the SM, although it does appear at the one-loop level. One immediate 
advantage of the $\gamma e\to W\nu$ process over, \eg, 
$e^+e^-\to W^+W^-$ is that the $\gamma WW$ vertex can be trivially isolated 
from the corresponding ones for the $ZWW$ vertex, thus allowing us to probe 
this particular vertex in a model-independent fashion. To set the notation for 
what follows, the $\gamma WW$ and $\gamma ZZ$ anomalous couplings are 
denoted by $\Delta \kappa$, $\lambda$  and $h_{3,4}^0${\cite {rev}}, 
respectively. We will assume that the $\gamma WW$ and $\gamma ZZ$ 
anomalous couplings are unrelated; the details of our analysis can 
be found in Ref.{\cite {old}}.

\section{Analysis}

The use of both polarized electron and photon beams allows one to construct a 
polarization asymmetry, $A_{pol}$. In general the $\gamma e \to W\nu ,Ze$ 
(differential or total) cross sections can be written schematically 
as $\sigma=(1+A_0P)\sigma_{un}+\xi(P+A_0)\sigma_{pol}$, where 
$P$ is the electron's polarization($>0$ for left-handed beams), 
$-1\leq \xi \leq 1$ is the Stoke's parameter for the circularly polarized 
photon, and $A_0$ describes the electron's coupling to the relevant gauge 
boson[$A_0=2va/(v^2+a^2)=1$ for $W$'s and $\simeq 0.145$ for $Z$'s, with $v,a$ 
being the vector and axial-vector coupling of the electron]. 
$\sigma_{pol}(\sigma_{un})$ 
represents the polarization (in)dependent contribution to the cross section, 
both of which are functions of only a single dimensionless variable at the 
tree level 
after angular integration, \ie, $x=y^2=s_{\gamma e}/M_{W,Z}^2$,  
where $\sqrt {s_{\gamma e}}$ is the $\gamma -e$ center of mass energy. 
Taking the ratio of the $\xi$-dependent to $\xi$ independent terms in $\sigma$ 
gives us the asymmetry $A_{pol}$.

\vspace*{-0.5cm}
\nn
\begin{figure}[htbp]
\centerline{
\psfig{figure=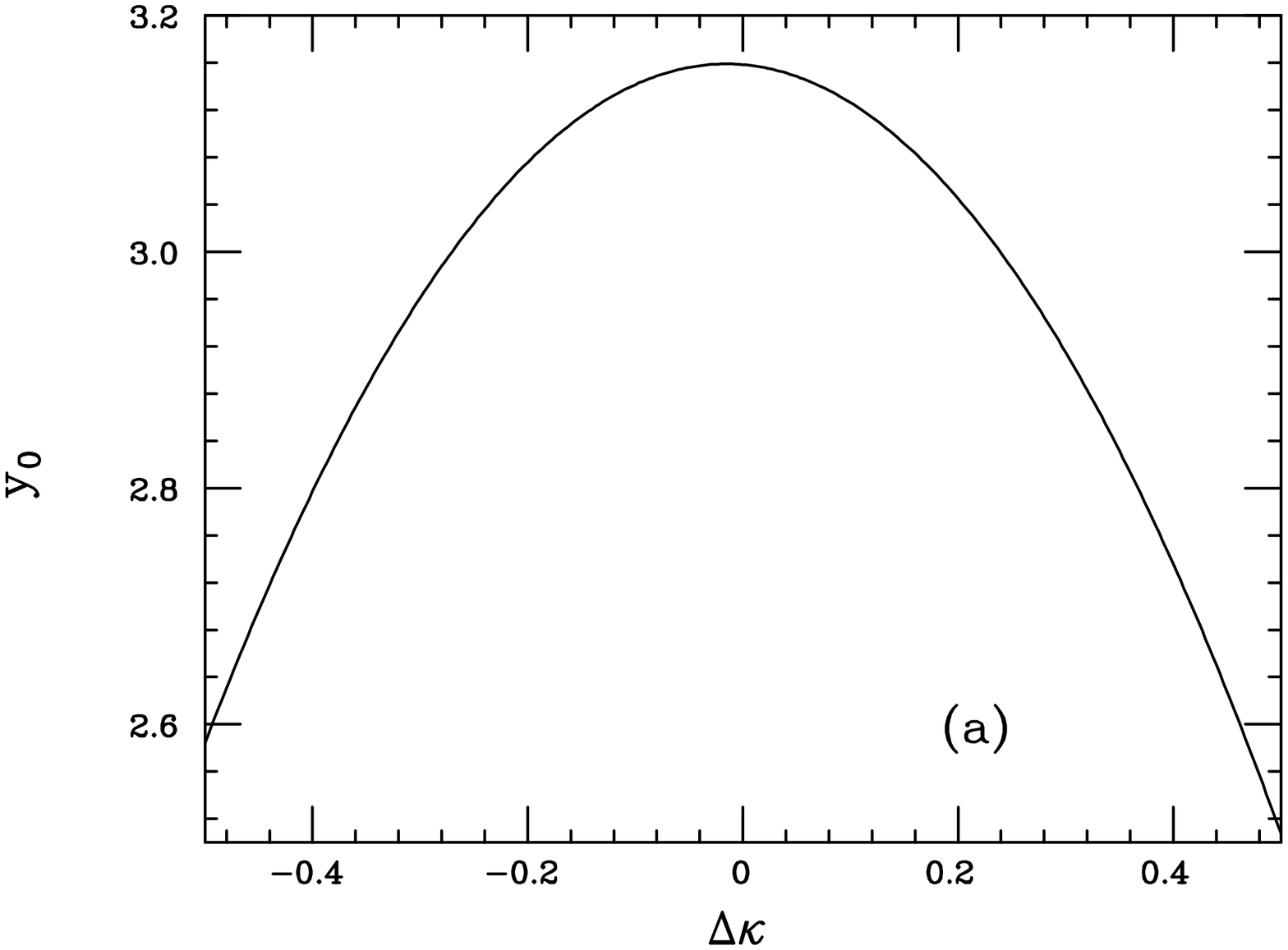,height=9.1cm,width=9.1cm,angle=90}
\hspace*{-5mm}
\psfig{figure=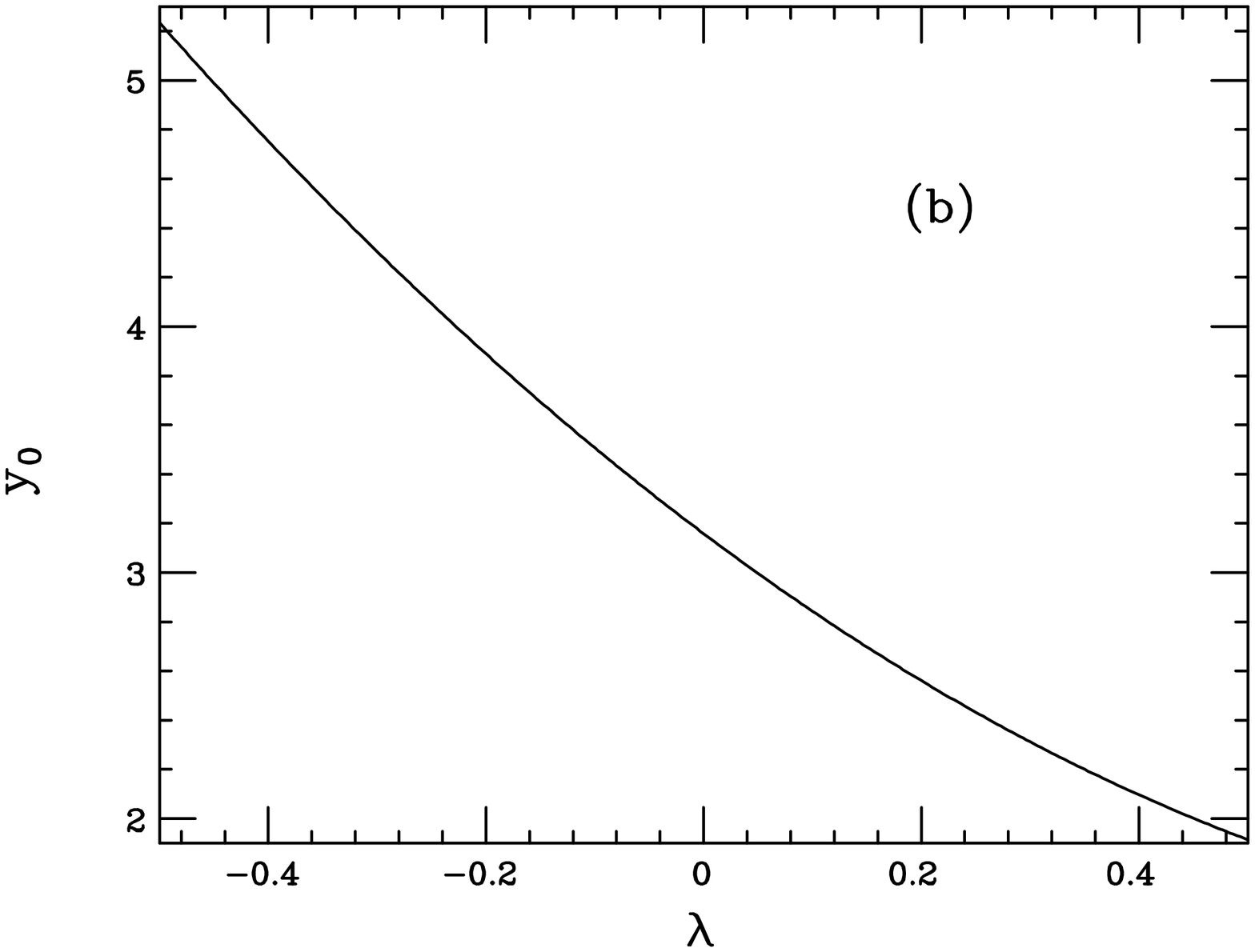,height=9.1cm,width=9.1cm,angle=90}}
\vspace*{-1cm}
\caption{\small Separate $\dk$ and $\lambda$ dependence of the
value of $y_0$, the zero position for the process $\gamma e \to W\nu$.}
\end{figure}
\vspace*{0.4mm}

One reason to believe {\it a priori} that $A_{pol}$, or $\sigma_{pol}$ itself,  
might be sensitive to modifications in the TGVs due to the presence of the 
anomalous couplings is the Drell-Hearn Gerasimov(DHG) Sum Rule{\cite {dhg}}.  
In its $\gamma e \to W \nu, Ze$ manifestation, the DHG sum rule implies that
\begin{equation}
\int_{1}^{\infty} {\sigma_{pol}(x)\over {x}} dx = 0 \,,
\end{equation}
for the tree level SM cross section when the couplings of all the 
particles involved in the process are `canonical', \ie, gauge invariant. 
That this integral is zero results from ($i$) the fact that 
$\sigma_{pol}$ is well 
behaved at large $x$ and ($ii$) a delicate cancellation occurs 
between the two 
regions where the integrand takes on opposite signs. This observation is 
directly correlated with the existence of a single 
value of $x$(or $y$) where $\sigma_{pol}$(and, hence, $A_{pol}$) vanishes. 
For the $W(Z)$ case this asymmetry `zero' occurs at 
$\sqrt {s_{\gamma e}}\simeq 254(150)$ GeV, both 
of which are easily accessible at the NLC. As we will see, the 
inclusion of anomalous 
couplings not only moves the position of the zero but also forces the 
integral to become non-vanishing and, in most cases, {\it infinite}. 
Unfortunately, since we cannot go to infinite energies we cannot test the DHG 
Sum Rule directly. In the 
$W$ case, the zero position, $y_0$, is found to be far more sensitive to 
modifications in the TGVs than in the $Z$ case. The zero position as a 
function of $\Delta \kappa$ and $\lambda$ for the $\gamma e\to W\nu$ process 
is shown in Fig.1 whereas the corresponding $Z$ case is shown in Fig.2. In 
either situation, the position of the zero 
{\it alone} does not offer great sensitivity to the existence of anomalous 
couplings.(See Ref. 2.)

\vspace*{-0.5cm}
\nn
\begin{figure}[htbp]
\centerline{
\psfig{figure=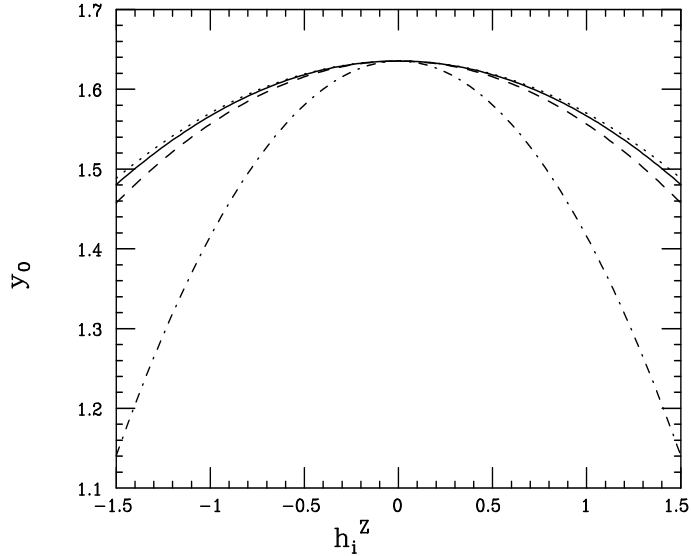,height=9.1cm,width=11cm,angle=-90}}
\vspace*{-1cm}
\caption{\small Position of the SM polarization asymmetry zero in 
$\gamma e \to Ze$ as a function of $h_{3,4}^0$ for $P=90\%$ with a $10^\circ$ 
angular cut. The dotted(dashed, 
dash-dotted, solid) curve corresponds to the case $h_4^0=0$($h_3^0=0$, 
$h_3^0=h_4^0$, $h_3^0=-h_4^0$).}
\end{figure}
\vspace*{0.4mm}

Our analysis begins by examining the energy, \ie, $y$ dependence of $A_{pol}$ 
for the two processes of interest; we consider the $W$ case first. For a 
500(1000) GeV collider, we see that only the range $1\leq y\leq 5.4(10.4)$ is 
kinematically accessible since the laser photon energy 
maximum is $\simeq 0.84E_e$. Since we are interested in bounds on the 
anomalous couplings, we will assume that the SM is valid and generate a set 
of binned $A_{pol}$ data samples via Monte Carlo taking 
only the statistical errors into account. We further assume that 
the electrons are 
90$\%$ left-handed polarized as right-handed electrons do not interact 
through the $W$ charged current couplings. Our bin width will be assumed to be 
$\Delta y=$0.1 or 0.2. 
We then fit the resulting distribution to 
the $\Delta \kappa$- and $\lambda$-dependent functional form of $A_{pol}(y)$ 
and subsequently 
extract the 95$\%$ CL allowed ranges for the anomalous couplings. The results 
of this procedure are shown in Fig. 3, where we see that reasonable 
constraints are obtained although only a single observable has been used in 
the fit. 

\nn
\vspace*{0.1mm}
\hspace*{-0.5cm}
\begin{figure}[htbp]
\centerline{\psfig{figure=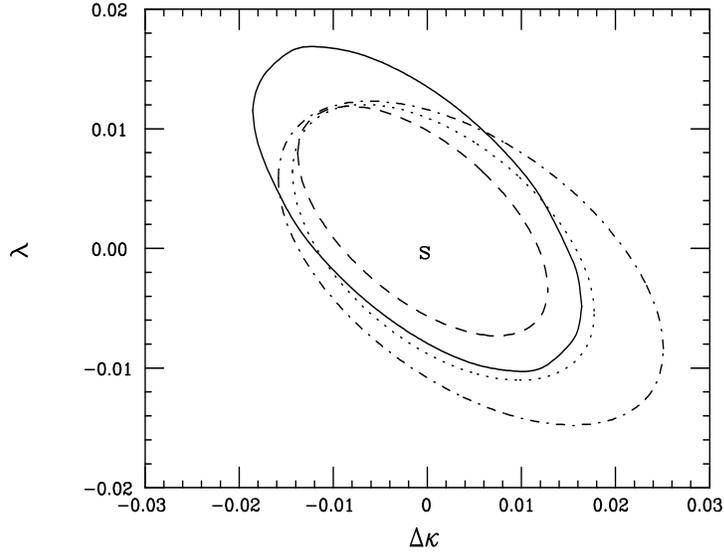,height=9.1cm,width=11cm,angle=90}}
\vspace*{-1.0cm}
\caption{\small 95 $\%$ CL bounds on the $W$ anomalous couplings from the 
polarization asymmetry. The 
solid(dashed, dash-dotted) curves are for a 500 GeV NLC 
assuming complete $y$ coverage using 22(22, 44) bins and an integrated 
luminosity per bin of 2.5(5, 1.25)$fb^{-1}$, respectively. The corresponding 
bins widths are $\Delta y=$0.2(0.2, 0.1). The dotted curve 
corresponds to a 1 TeV NLC using 47 $\Delta y=0.2$ bins with 2.5 $fb^{-1}$/bin. 
`s' labels the SM prediction.}
\end{figure}

Clearly, to obtain stronger limits we need to make a combined fit with other 
observables, such as the energy dependence of the total cross section, the 
$W$ angular distribution, or the $W$ polarization. As an example we 
show in Fig. 4 that the size of 
the $95\%$ CL allowed region shrinks drastically in the 1 TeV case when the 
cross section data is included in a simultaneous fit together 
with the polarization 
asymmetry. As is well known, the cross section is highly sensitive to 
$\Delta \kappa$ and thus the allowed region is highly compressed in that 
direction. We find that $\Delta \kappa$ is bounded to the range 
$-1.45\cdot 10^{-3}\leq \Delta \kappa \leq 0.36\cdot 10^{-3}$ while the 
allowed $\lambda$ range is still rather large.
The addition of the angular distribution and $W$ polarization data 
to the fit is expected 
to reduce the size of this allowed region even further.

\nn
\vspace*{0.1mm}
\hspace*{-0.5cm}
\begin{figure}[htbp]
\centerline{\psfig{figure=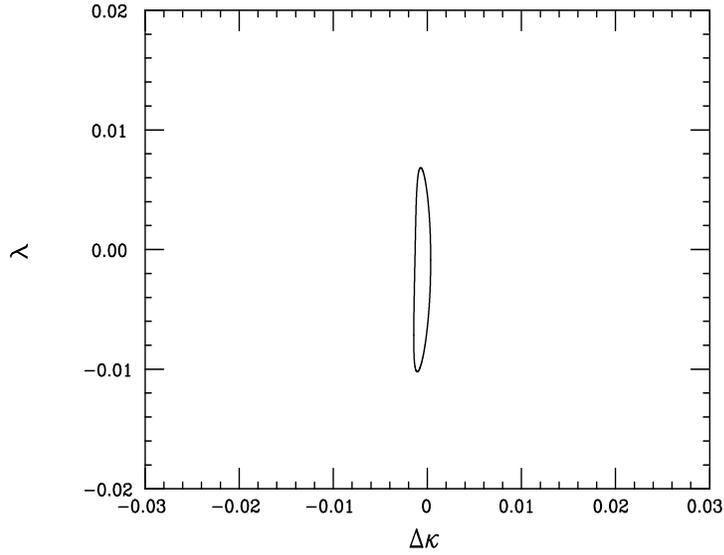,height=9.1cm,width=11cm,angle=-90}}
\vspace*{-1.0cm}
\caption{\small Same as the previous figure for a 1 TeV NLC but now 
combined with the cross section data in a simultaneous fit. Only 
statistical errors are included.}
\end{figure}

With these thoughts in mind, in the $Z$ case we will follow a similar approach 
but we 
will simultaneously fit both the energy dependence of $A_{pol}$ as well as 
that of 
the total cross section. (Later, we will also include the $Z$ boson's angular 
distribution 
into the fit.) In this $Z$ analysis we make a $10^{\circ}$ angular cut on the 
outgoing electron and keep a finite form factor scale, $\Lambda=1.5$ TeV, so 
that we may more readily compare with other existing analyses. (The angular 
cut also gives us a finite cross section in the massless electron limit; 
this cut is not required in the case of the $W$ production process.) We again  
assume that $P=90\%$ so that this analysis can take place simultaneously with 
that for the $W$. The accessible $y$ ranges are now 
$1\leq y \leq 4.6(9.4)$ for a 500(1000) GeV collider. Fig.5 shows our results 
for the 500 GeV NLC while Fig.6 shows the corresponding 1 TeV case. For a 
given energy and fixed total integrated luminosity we learn from these figures 
that it is best to 
take as much data as possible at the highest possible values of $y$. 
Generally, one finds that increased sensitivity to the existence of anomalous 
couplings occurs at the highest possible collision energies.

Even these anomalous coupling bounds can be significantly 
improved by including the $Z$ boson angular information in 
the fit. To be concrete we examine the case of a 1 TeV NLC with 
16.8$fb^{-1}$/bin of integrated luminosity taken in the last 10 $\Delta y$ 
bins(corresponding to the dash-dotted curve in Fig.6). Deconvoluting the 
angular integration and performing instead the integration over the 10 
$\Delta y$ bins we obtain the energy-averaged angular distribution. Placing 
this distribution into 10 (almost) equal sized $cos \theta$ bins while still 
employing our $10^\circ$ cut, we can use this 
additional data in performing our overall simultaneous $\chi^2$ fit. The 
result of doing this is shown in Fig.7 together with 
the anticipated result from the LHC using the $Z\gamma$ production mode. Note 
that the additional angular distribution data has reduced the size of the 
$95\%$ CL allowed region by almost a factor of two. 
Clearly both machines are complementary in their abilities to probe small 
values of the $\gamma ZZ$ anomalous couplings. If the NLC and LHC results were 
to be combined, an exceptionally small allowed region would remain. 
The NLC results themselves may be further improved by considering 
measurements of the polarization of the final state $Z$ as well as 
by an examination of, \eg, the complementary $e^+e^- \to Z\gamma$ process;  
such studies are currently underway{\cite {joa}}.

\vspace*{-0.5cm}
\nn
\begin{figure}[htbp]
\centerline{
\psfig{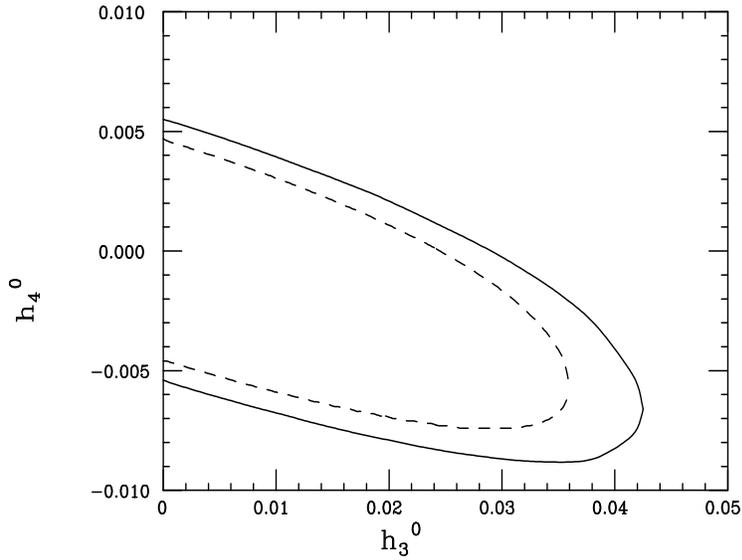}}
\vspace*{-1cm}
\caption{\small $95\%$CL allowed region for the anomalous coupling parameters 
$h_3^0$ and $h_4^0$ from a combined fit to the energy dependencies of the total 
cross section and polarization asymmetry at a 500 GeV NLC assuming $P=90\%$ 
and an integrated luminosity of $3(6)fb^{-1}$/bin corresponding to the solid
(dashed) curve. 18 bins of width $\Delta y$=0.2 were chosen to cover the $y$ 
range $1\leq y \leq 4.6$. The corresponding bounds for negative values of 
$h_3^Z$ are obtainable by remembering the invariance of the polarization 
dependent cross section under the reflection $h_{3,4}^0\to -h_{3,4}^0$.}
\end{figure}
\vspace*{0.4mm}

\section{Discussion and Conclusions}

The collision of polarized electron and photon beams at the NLC offers an 
exciting opportunity to probe for anomalous gauge couplings of both the $W$ 
and the $Z$ through the use 
of the polarization asymmetry. In the case of $\gamma e \to W\nu$ we can 
cleanly isolate the $\gamma WW$ vertex in a model independent fashion. When 
combined with other observables, extraordinary sensitivities to such 
couplings for $W$'s are achievable at the NLC in the $\gamma e$ mode. These are 
found to be quite complementary to those obtainable in $e^+e^-$ collisions.
In the case of the 
$\gamma ZZ$ anomalous couplings, we obtained constraints comparable to those 
which can be obtained at the LHC.

\Acknowledgements

The author would like to thank S. J. Brodsky, I. Schmidt, J.L. Hewett, and   
S. Godfrey for discussions related to this work.

%
%%%%%%%%%%%%%%%%%%--- References
%%%%%%%%%%%%%%%%%%%%%%%%%%%%%%%%%%%%%%%%%%%%%%%%%%%%%%%
\def\MPL #1 #2 #3 {Mod.~Phys.~Lett.~{\bf#1},\ #2 (#3)}
\def\NPB #1 #2 #3 {Nucl.~Phys.~{\bf#1},\ #2 (#3)}
\def\PLB #1 #2 #3 {Phys.~Lett.~{\bf#1},\ #2 (#3)}
\def\PR #1 #2 #3 {Phys.~Rep.~{\bf#1},\ #2 (#3)}
\def\PRD #1 #2 #3 {Phys.~Rev.~{\bf#1},\ #2 (#3)}
\def\PRL #1 #2 #3 {Phys.~Rev.~Lett.~{\bf#1},\ #2 (#3)}
\def\RMP #1 #2 #3 {Rev.~Mod.~Phys.~{\bf#1},\ #2 (#3)}
\def\ZP #1 #2 #3 {Z.~Phys.~{\bf#1},\ #2 (#3)}
\def\IJMP #1 #2 #3 {Int.~J.~Mod.~Phys.~{\bf#1},\ #2 (#3)}
\vspace*{-0.5cm}
\nn
\begin{figure}[htbp]
\centerline{
\psfig{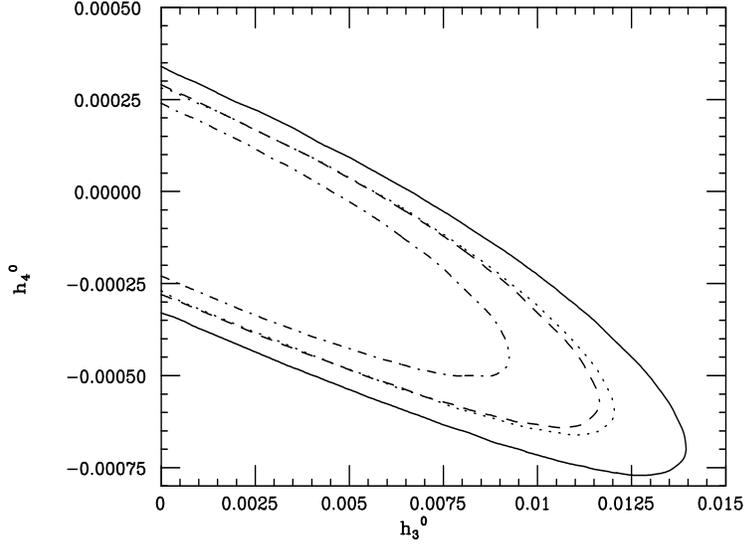}}
\vspace*{-1cm}
\caption{\small Same as Fig. 5 but for a 1 TeV NLC. The solid(dashed) curve 
corresponds to a luminosity of $4(8)fb^{-1}$/bin for 42 bins of width 
$\Delta y$=0.2 which covered the range $1\leq y \leq 9.4$. The dotted curve 
corresponds to a luminosity of $8fb^{-1}$/bin but only for the last 21 bins. 
The dash-dotted curve corresponds to the case of $16.8fb^{-1}$/bin in only the 
last 10 bins.}
\end{figure}
\vspace*{0.4mm}
\vspace*{-0.5cm}
\nn
\begin{figure}[htbp]
\centerline{
\psfig{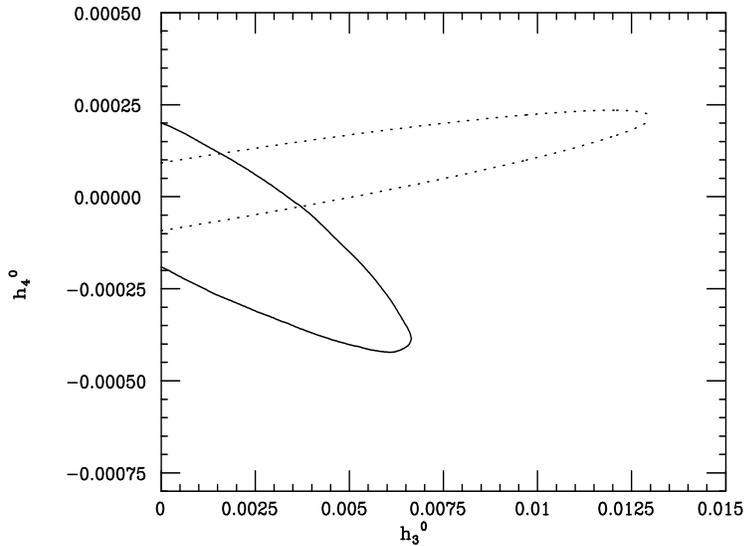}}
\vspace*{-1cm}
\caption{\small The solid curve is the same as dash-dotted curve 
in Fig. 6, but now including in the fit 
the $Z$ boson angular distribution obtained from the highest 
10 bins in energy. The 
corresponding result for the 14 TeV LHC with 100$fb^{-1}$ of integrated 
luminosity from the process $pp\to Z\gamma+X$ is shown as the dotted curve.}
\end{figure}
\vspace*{0.4mm}

\end{document}